\documentstyle[twoside,fleqn,espcrc2,epsf]{article}

\newcommand{\AmS}{{\protect\the\textfont2
  A\kern-.1667em\lower.5ex\hbox{M}\kern-.125emS}}

\title{ Gluino-Mediated Rare B Decays 
 \thanks{Based on an invited talk given at the International 
\mbox{Euroconference} 
  on Quantum Chromodynamics, 
July 1999, Montpellier, France, to appear in Nucl. Phys. B Suppl.}
\hspace{3cm} {\small\it CERN-TH/99-342,   MPI/PhT-99-43}  } 
\author{Tobias Hurth\address{CERN, Theory Division,\\
CH-1211 Geneva 23, Switzerland}\address{Max-Planck-Institute for Physics,
  Werner-Heisenberg-Institute,\\
F\"ohringer Ring 6, D-80805 Munich, Germany}}%

\begin{document}

\begin{abstract}
We discuss  the gluino-induced contribution
to rare B decays in supersymmetric
frameworks with generic sources of flavour change.
\end{abstract}

\maketitle

\section{Introduction}

Apart from the low-energy regime of the strong interaction,
flavour physics is the least tested part of the 
SM. This is reflected in the rather large error bars of 
several flavour parameters such as the mixing parameters
at the twenty percent level \cite{Parodi}.
%which has to be compared with 
%errors smaller than one percent in electroweak precision experiments.
However, the experimental situation concerning $B$ physics 
will drastically change in the near future.  
There are several $B$ physics experiments successfully
running at the moment. In the upcoming years new facilities will start
to explore  $B$ physics with increasing 
sensitivity and within   different  
experimental settings \cite{Artuso}.

The $b$ quark system 
is an ideal laboratory for studying flavour physics. 
Hadrons containing a $b$ quark 
are the heaviest hadrons experimentally accessible. 
Since the mass of the $b$ quark is much larger than the QCD scale, 
the long-range strong interactions are expected to be comparably
small and are well under control thanks to the  heavy quark 
expansion~\cite{Isgur,Neuberta}.  

Of particular interest are the so-called rare $B$ decays, which are 
flavour  changing neutral current processes (FCNC) which
vanish at the tree level of the SM. Thus, they 
are rather sensitive probes for physics beyond the SM \cite{Susya,a}.
%The $B \to X_s \gamma$ decay, for example, plays 
%already a very important role in  restricting the parameter space of 
%extensions  of the SM like the minimal supersymmetric standard
%model (MSSM) \cite{Susya,SusySusya,a} in spite of the fact that the 
%accuracy of the
%experimental data on  $B \to X_s \gamma$   used in such analyses 
%is less  than $30 \%$. 

One of the main difficulties in analysing rare $B$ decays is the 
calculation of short-distance QCD effects. These radiative corrections 
lead to a tremendous rate enhancement.
The QCD radiative corrections  bring in large logarithms of the form 
$\alpha_s^n(m_b) \, \log^m(m_b/M)$,
where $M$ is the top quark or the W mass and $m \le n$ (with $n=0,1,2,...$).
%This is a natural  feature in any process where two different mass scales
%are present.
They have to get resummed at least to 
leading-log (LL) precision ($m=n$).

Within the SM the accuracy
in  the domina- ting perturbative contribution to 
$B \rightarrow X_s \gamma$ 
was recently  improved to NLL
precision.
This was a joint effort of many different groups 
 \cite{AG91}. The theoretical error of the  previous 
leading-log (LL) result was substantially reduced 
to $\pm 10\%$ and the central value of the partonic
decay rate increased by about $20\%$.

Supersymmetric extensions of the SM have become the most popular 
framework of new theoretical structures at higher scales,
much below the Planck scale. 
The precise mechanism of the necessary supersymmetry breaking
is unknown \cite{question}. A reasonable approach to this problem 
is the inclusion of the most general 
soft breaking term  consistent with the SM gauge symmetries in the
so-called unconstrained minimal supersymmetric standard model (MSSM). 
This leads to a proli- feration of free  parameters in the theory. 

A global fit to electroweak precision measurements within supersymmetric
models shows that 
if the superpartner spectrum becomes light the fit to the data results
in typically larger values of $\chi^2$ compared with the SM.  
%Significant portions of the parameter spaces of popular supersymmetry
%breaking scenarios are excluded. 
Supersymmetric \mbox{models}, however, 
can always avoid serious constraints from data because the supersymmetric 
contributions decouple \cite{Damien}.

In the MSSM there are two kinds of new contributions to FCNC processes.
The first class results from flavour mixing in the sfermion mass matrices
\cite{DNW}. Moreover, one has CKM-induced contributions from charged Higgs
boson and chargino exchanges (see \cite{miksum}). 
This leads to the well-known supersymmetric flavour problem: the severe
experimental constraints on flavour violations have no direct explanation 
in the structure of the unconstrained MSSM.  
Clearly, the origin of flavour violation is a model-dependent issue
and  is based 
on the relation of the dynamics of flavour and 
the mechanism of supersymmetry breaking. 
Keeping in mind our current phenomenological knowledge about 
supersymmetry, it 
is suggestive to 
perform  a model-independent analysis of flavour changing phenomena.  
Such an analysis  provides important hints on the more 
fundamental theory of soft supersymmetry  breaking.

\section{Gluino Contribution to $B \rightarrow X_s \gamma$}
 
Among inclusive rare $B$ decays, the $B \rightarrow X_s \gamma$ mode
is the most prominent because it is the only decay mode in this class 
that is already measured \cite{CLEOneu,ALEPH}.
Many papers are devoted to studying the $B \rightarrow X_s \gamma$ decay  
and similar decays within the MSSM.
However, in most of these analyses, the contributions of
supersymmetry were not investigated  with the systematics of the SM
 calculations. 
In \cite{guidice} it was shown, 
that in a specific supersymmetric scenario NLL contributions are 
important and lead to a significant reduction of the stop-chargino 
mass region where the supersymmetric contribution has
a large destructive interference with the charged-Higgs boson 
contribution. 
%The NLL two-loop matching contributions with gluons were 
%recently presented in \cite{mikolaj99}.
It is expected that the complete NLL calculation drastically decreases 
the scale dependence and, thus,  the \mbox{theoretical} error.  
The NLL analysis 
is also a necessary check of the validity of the perturbative ansatz
(see \cite{BG}).
The NLL calculations in \cite{guidice} and 
also in \cite{mikolaj99} are worked out in the
heavy gluino case. In the analysis \cite{complete} reported here, 
the gluino-mediated decay $B \rightarrow X_s \gamma$
is discussed where the gluino is not assumed to be decoupled.

Previous work on the gluino contribution 
\cite{DNW,HKT,GGMS}
did not include LL or NLL QCD corrections,
and  gluino exchanges were
treated in the so-called mass insertion approximation (MIA) 
%\cite{MASSINSERTION} 
only, 
where the off-diagonal squark mass matrix elements are taken to be small 
and their higher
powers neglected.
In our analysis we explore the limits of the MIA.
Furthermore, we analyse the sensitivity  of the bounds 
on the  sfermion mass matrices   to radiative QCD corrections.

Within the SM, there is one 
coupling constant, $G_F$, relevant to the $b \rightarrow s \gamma$
decay. There is also one flavour violation parameter only, 
namely the product 
of two CKM matrices. All the loops \mbox{giving} the logarithms  
are due to gluons, which imply a factor of $\alpha_s$.
The corrections can then be classified according to: 
\begin{itemize}
\item (LL), $G_F (\alpha_s Log)^N$, 
\item (NLL),  $G_F \alpha_s (\alpha_s Log)^N$.
\end{itemize}
Thus, the above ordering also reflects the actual size of the
contributions to $b \rightarrow s \gamma$. 

The corresponding analysis of QCD corrections  in the MSSM is much more 
complicated. 
The MSSM has several couplings
relevant to this decay and there are several 
flavour changing parameters. Thus, a formal LL term might have a 
small coupling
while a NLL contribution is multiplied with a large one.
Moreover, the couplings generally depend on the parameters,
and the results should be applicable for large domains on
the parameters.

Another complication in supersymmetric theo- ries is the occurrence of flavour 
violations  such as gluino exchanges
(through the gluino-quark-squark coupling)
where additional factors $\alpha_s$ 
are induced.
They lead to magnetic penguin operators where the Wilson 
coefficients naturally contain factors of $\alpha_s$. 
Moreover, these contributions induce  magnetic operators where  
the (small) factor $m_b$ is replaced by the gluino mass. Clearly this  
contribution is expected to be
dominating.  
The gluino-induced contributions to the decay amplitude for $b \to s \gamma$
are of the following form:
\begin{itemize}
\item (LL),  $\alpha_s \, (\alpha_s Log)^N$
\item (NLL), $\alpha_s \, \alpha_s (\alpha_s Log)^N$
\end{itemize}

In the matching calculation, 
all factors $\alpha_s$  regardless of their source should get 
expressed in terms of the $\alpha_s$ running with five flavours.
However, non-decoupling effects through
violations of the supersymmetric equivalence between gauge bosons and 
corresponding gaugino couplings have to be taken 
into account at the NLL level.

Furthermore, one finds that  
gluino-squark boxes induce new scalar and tensorial 
four-Fermi operators, which are shown to 
mix into the magnetic operators without gluons.  
%If these four Fermi operators are defined with an explicit
%factor $\alpha_s^2$,
%the anomalous dimension matrix has the canonical expansion
%in $\alpha_s$.
On the other hand, the vectorial four-Fermi operators 
mix only with an
additional gluon into magnetic ones. 
Thus, they will contribute at the next-to-leading order only.
However, from the numerical point
of view the contributions of the vectorial operators (although NLL) are
not necessarily suppressed w.r.t the new four-Fermi contributions;
this is due to the expectation 
that the flavour-violation parameters
present in the Wilson coefficients of the new operators are expected
to be much smaller (or much more stringently constrained)
than the corresponding ones in the coefficients of the vectorial 
operators. This is one of the most important reasons why a complete NLL
order calculation should be performed. 

The mixed graphs, containing a $W$, a gluino and a squark, 
are proportional to $G_F \alpha_s$. They give rise only to corrections
of the SM operators at the NLL level.
There are also penguin contributions with two gluino lines
in the NLL matching.

The current  discussion is restricted to the $W$ or 
gluino-mediated flavour changes and does not consider contributions 
with other Susy particles such as chargino, charged Higgs or neutralino. 
Clearly, analogous phenomena occur  in those contributions.

To understand the sources of flavour violation that may be present in
supersymmetric models in addition to those enclosed in the CKM matrix,
one has to consider the contributions to the mass
matrix of a squark of flavour $f$: 
\begin{equation}
{\cal M}_{f}^2 =  
\left( \begin{array}{cc}
  m^2_{f,LL}   & m^2_{f,LR} \\
  m^2_{f,RL}  &  m^2_{f,RR}                 
 \end{array} \right) +
\label{squarku}
\nonumber
\end{equation}
\begin{equation}
 + \left( \begin{array}{cc}
  F_{f,LL} +D_{f,LL} &  F_{f,LR} \\
 F_{f,RL} & F_{f,RR} +D_{f,RR}                
 \end{array} \right) 
\nonumber
\label{squarku2}
\end{equation}

In the super  CKM basis where the quark mass matrix is diagonal 
and the squarks are rotated in parallel to their superpartners,
the $F$ terms  from the superpotential and the $D$ terms 
in the $6 \times 6$
mass matrices ${\cal M}^2_f$ turn out to be diagonal 
$3 \times 3$ submatrices. This is in general not true
for the additional terms (\ref{squarku}) from  the soft 
supersymmetry breaking potential. 
Because all neutral gaugino couplings are flavour diagonal
in the super CKM basis, the gluino contributions to the
decay width of $b \to s \gamma$ are induced by the off-diagonal
elements of the soft terms 
$m^2_{f,LL}$, $m^2_{f,RR}$, $m^2_{f,RL}$.

\section{Numerical Results}
\label{constraints}

We show a few features of our numerical results
based on a complete LL calculation. More details  
of the analysis can be found in \cite{complete}.
%We show  the interplay between the standard
%model and the gluino contribution to the decay $b \to s \gamma$.
The size of the gluino contribution crucially depends on the soft
terms in the squark mass matrix ${\cal M}_{\tilde{D}}^2$ and to a lesser
extent on those in ${\cal M}_{\tilde{U}}^2$. 
In the following, we take all the diagonal entries in the soft matrices
$m^2_{\tilde{Q},LL}$,
$m^2_{\tilde{D},RR}$,
$m^2_{\tilde{U},RR}$,
to be equal;
their common mass is denoted by $m_{\tilde{q}}$ and set to the value $500$ GeV.
First, the matrix element 
$m^2_{\tilde{D},LR;23}$ 
is varied. All other entries in the soft mass terms are put to zero.
Following the notation of \cite{GGMS},
we define 
\begin{equation} 
\delta_{LR;23} = m^2_{\tilde{D},LR;23}/m^2_{\tilde{q}} \quad 
\mbox{and} \quad x=m^2_{\tilde{g}}/m^2_{\tilde{q}} \, ,
\label{deltadef}
\end{equation}
where $m_{\tilde{g}}$
is the gluino mass.
%The parameters in the supersymmetric mass 
%terms are chosen to be: $\mu=0$, $\tan \beta =2$, $m_b =3$ GeV, 
%$m_t = 175$ GeV, 
%$m_Z=91.18$ GeV, and $\sin^2 \theta_W = 0.2316$.
%\begin{figure}[th]
\begin{figure}[p]
\begin{center}
\leavevmode
\epsfxsize= 6.0 truecm
\epsfbox[18 167 580 580]{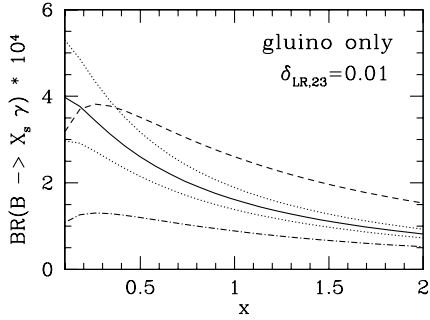}
\end{center}
\caption[f1]{Gluino-induced branching ratio $BR(B \to X_s \gamma)$ 
as a function of $x= m^2_{\tilde{g}}/m^2_{\tilde{q}}$,
when only $\delta_{LR,23}$ is non-vanishing (see text).}
\label{sizeqcd23ll}
\end{figure}
\begin{figure}[p]
\begin{center}
\leavevmode
\epsfxsize= 6.0 truecm
\epsfbox[18 167 580 580]{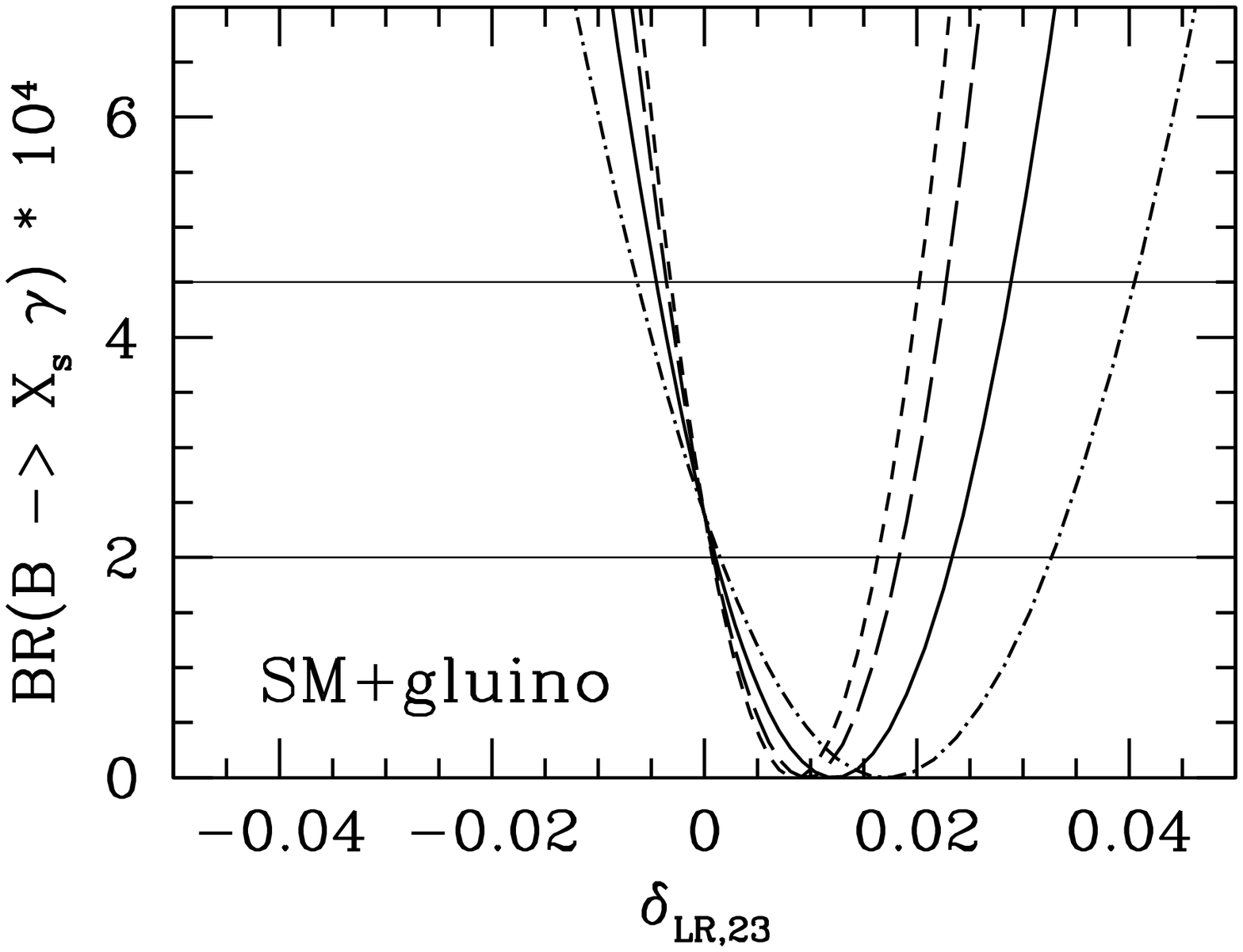}
\end{center}
\begin{center}
\leavevmode
\epsfxsize= 6.0 truecm
\epsfbox[18 167 580 580]{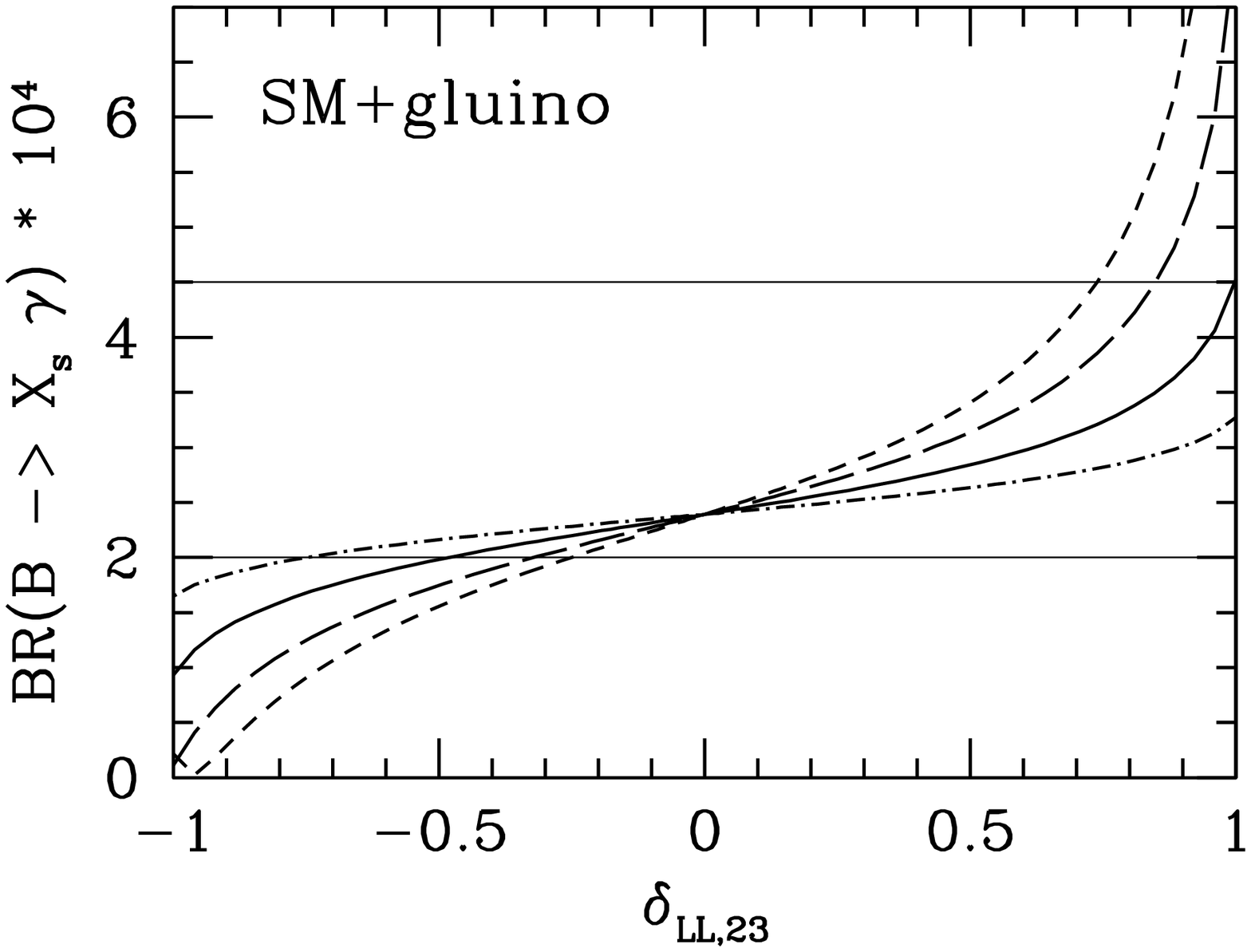}
\end{center}
\caption[f1]{$BR(B \to X_s \gamma)$  
with $x$ = 0.3 (short-dashed
 line), 0.5 (long-dashed line), 1 (solid line), 2 (dot-dashed line), see text.}
\label{glsm23ll}
\end{figure}
\begin{figure}[p]
\begin{center}
\leavevmode
\epsfxsize= 6.0 truecm
\epsfbox[18 167 580 580]{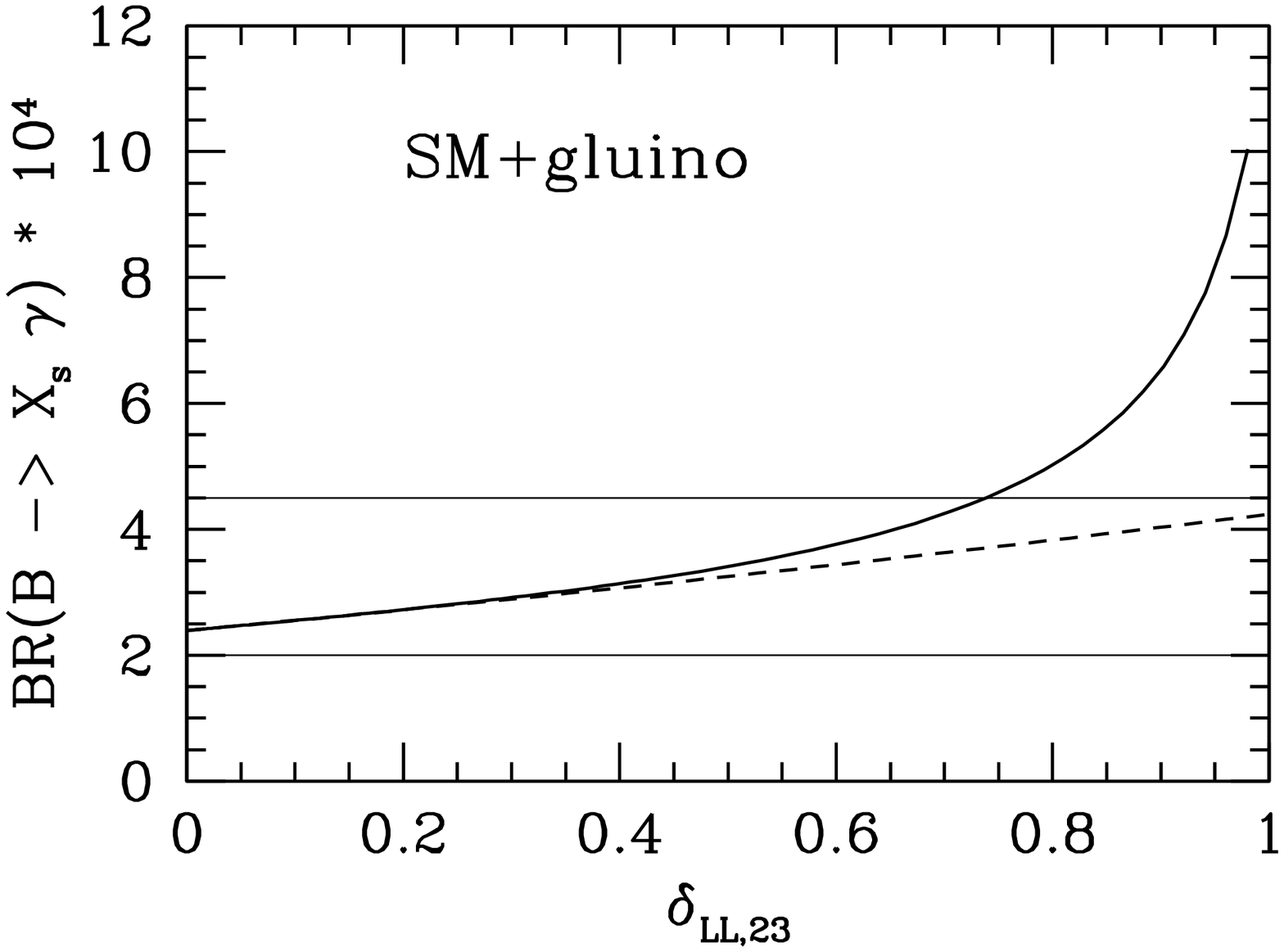}
\end{center}
\begin{center}
\leavevmode
\epsfxsize= 6.0 truecm
\epsfbox[18 167 580 580]{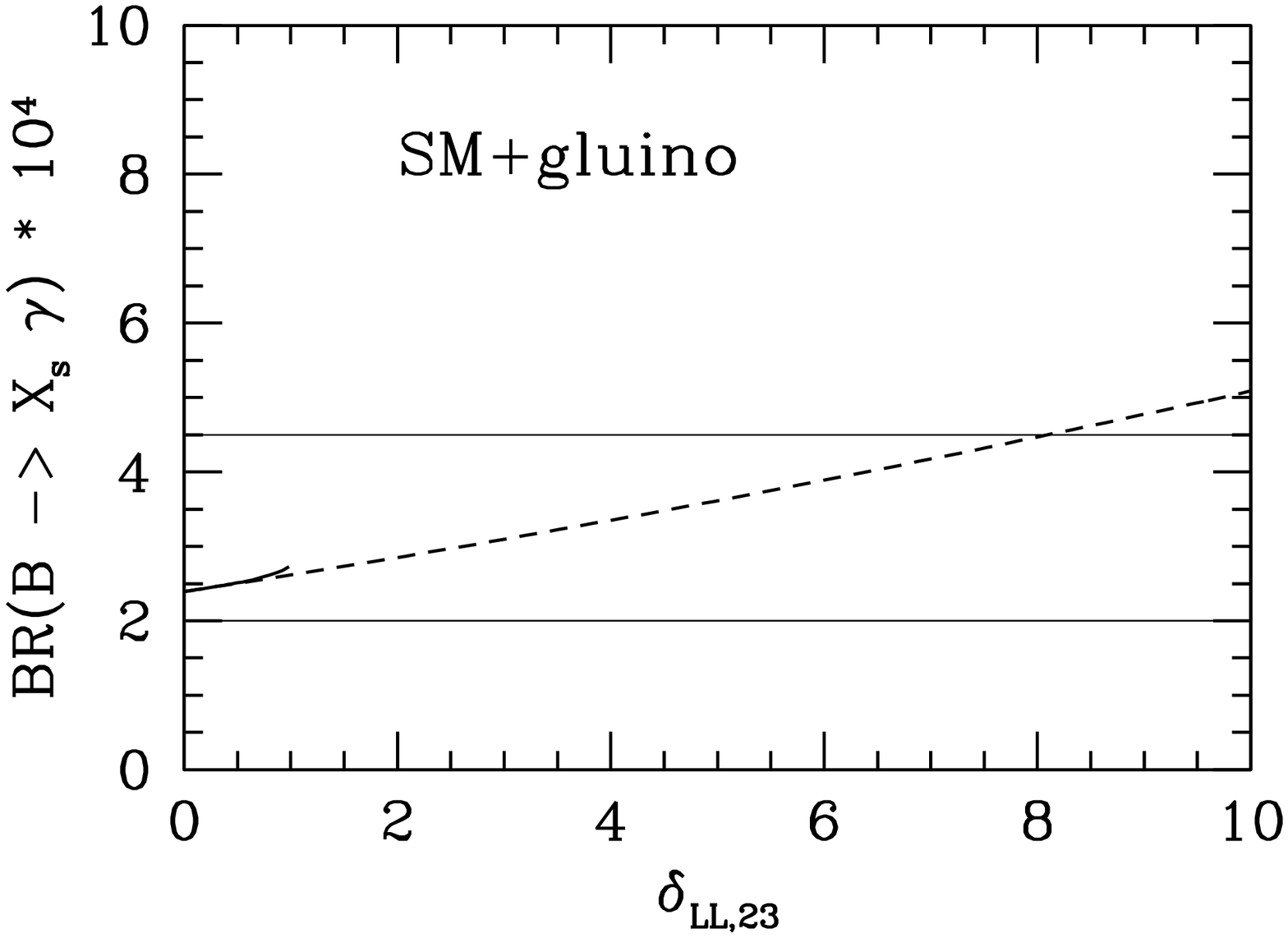}
\end{center}
\caption[f1]{Mass insertion approximation (dashed line) vs. exact result 
(solid line) as a function of $\delta_{LL;23}$ (see text).} 
\label{massins_ll03}
\end{figure}
\begin{figure}[p]
\begin{center}
\leavevmode
\epsfxsize= 6.0 truecm
\epsfbox[18 167 580 580]{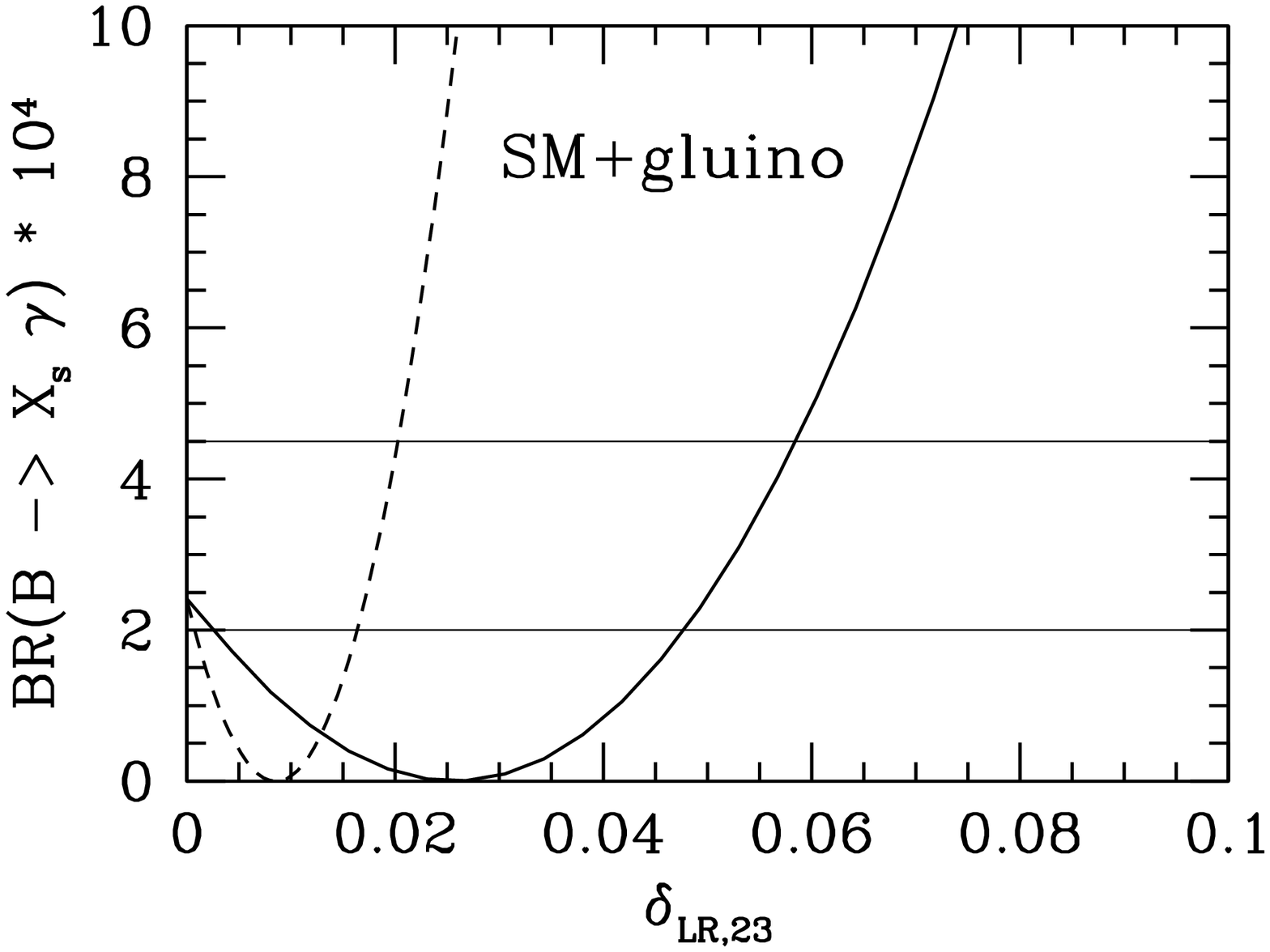}
\end{center}
\caption[f1]{$BR(B \to X_s \gamma)$ as a function of $\delta_{LR;23}$
including interference effects with a chain $\delta_{LL;23} \delta_{LR;33}$ in solid line 
(see text).}
\label{ll23lr33lr23}
\end{figure} 
%In the following, the SM contribution to $BR(B \to X_s \gamma)$ is, in
%general, added to the gluino contribution: possible constraints to the
%flavour violating sources in the squark sector should be extracted
%keeping into account that the SM contribution already successfully
%saturate the experimental result this branching ratio~\cite{EXPER}. 
%illustrate the impact of the LL QCD corrections on the gluino induced
%Hamiltonian. 
In Fig.~\ref{sizeqcd23ll}, 
the QCD-corrected
branching ratio is shown as  a function of
$x$ (solid lines), obtained when only $\delta_{LR,23}$ is vanishing 
($\delta_{LR,23}=0.01$).
Shown is also the range of variation
of the branching ratio, delimited by dotted lines, obtained when the
low-energy scale $\mu_b$ spans the interval $2.4$--$9.6\,$GeV. The
matching scale $\mu_W$ is here fixed to $m_W$. As can be seen, the
theoretical estimate of $BR(B \to X_s \gamma)$ is still largely
uncertain ($\sim \pm 25\%$). An extraction of bounds on the $\delta$
quantities more precise than just an order of magnitude, therefore,
would require the inclusion of next-to-leading logarithmic QCD
corrections. It should be noticed, however, that the inclusion of the
LL QCD corrections has already removed the large ambiguity on the
value to be assigned to the factor $\alpha_s(\mu)$ in the
gluino-induced operators. 
Before adding QCD corrections, it is not clear whether the explicit $\alpha_s$ 
factor should be taken at  some high scale $\mu_W$ or a some low scale 
$\mu_b$, the difference is a LL effect.
The corresponding values
for $BR(B \to X_s \gamma)$ for the two extreme choices of $\mu$ are
indicated in Fig.~\ref{sizeqcd23ll} by
the dot-dashed lines ($\mu=m_W$) and the dashed lines
($\mu=4.8\,$GeV). The branching ratio is then virtually unknown.  

In spite of the large uncertainties which the branching ratio 
$BR(B \to X_s \gamma)$ still has at the LL in QCD, it is possible
to extract indications on the size that the $\delta$-quantities 
may maximum acquire without inducing conflicts with the 
experimental measurements. As already noted in Ref.~\cite{GGMS}, 
the element $\delta_{LR,23}$ is certainly the flavour-violating
parameter most efficiently constrained. In Fig.~\ref{glsm23ll}               , 
the dependence of $BR(B \to X_s \gamma)$ is shown as a function 
of this parameter when this is the only flavour-violating source. 
The branching ratio is obtained by adding the SM and the 
gluino contribution obtained for different choices of $x$
for the fixed values 
$\mu_b = 4.8\,$GeV and $\mu_W = m_W$. 
The gluino  contribution interferes constructively with the SM 
for negative values of $\delta_{LR,23}$, which are then more 
sharply constrained than the positive values. Overall, this 
parameter cannot exceed the per cent level. 
Much weaker is the dependence on
$\delta_{LL,23}$ if this is the only off-diagonal
element in the down squark mass matrix. This dependence is illustrated
in Fig.~\ref{glsm23ll} for different choices of $x$. 
The induced
gluino contribution interferes constructively with the SM
contribution for positive $\delta_{LL,23}$.  
Notice that given the
large values of $\delta_{LL,23}$ allowed by the experimental
measurement, the MIA cannot be used in this
case to obtain a reliable estimate of $BR(B \to X_s \gamma)$, whereas
it is an excellent \mbox{approximation} of the complete calculation in the
case of $\delta_{LR,23}$.
In the upper frame of Fig.~\ref{massins_ll03},
we vary $\delta_{LL;23}$ for $x=0.3$.
The MIA and the exact result start to deviate considerably for
$x>0.4$, i.e. well within the experimental error band;
the exact result leads to more stringent bounds. An even more drastic
example is shown in the lower frame of Fig.~\ref{massins_ll03}, where
we increase $m_{\tilde{g}}$ in such a way 
that $x=4$, leaving all the other parameters
unchanged: for $\delta_{LL;23}>0.1$ 
the mass matrix  ${\cal M}_{\tilde{D}}^2$ 
has at least one negative eigenvalue.
This feature is of course completely missed in
the MIA.
One also has to consider
interference effects. In Fig.~\ref{ll23lr33lr23}
we show that the  additional contribution through
a chain $\delta_{LL;23} \delta_{LR;33}$ weakens
the bound on the parameter $\delta_{LR;23}$ significantly.
In the solid curve we put 
$\delta_{LL;23}=\delta_{LR;33}=
\sqrt{\delta_{LR;23}}$, while in the dashed curve 
$\delta_{LL;23}=\delta_{LR;33}=0$.
We have chosen again $x=0.3$.

Finally, we stress that a consistent precision analysis
of the bounds on the sfermion mass matrix 
should include a NLL calculation and also 
interference effects with the chargino contribution. \\

%{\tt\ttbs} 

\thanks 

The work reported here has been done in
collaboration with F. Borzumati, C. Greub 
and \mbox{D. Wyler}, which is gratefully acknowledged.

\end{document}